\documentclass[final,times,5p]{elsarticle}

\journal{Physics Letters B}

\usepackage{amssymb,amsmath}
\usepackage{mathrsfs}
\usepackage{textpos}
\usepackage{calc}
\usepackage{graphicx}
\usepackage{afterpage}
\usepackage{underscore}
\usepackage{dcolumn}
\usepackage{multirow}
\usepackage{srcltx}
\usepackage{color}
\usepackage[usenames,dvipsnames,svgnames,table]{xcolor}
\usepackage{xspace}

\newcommand{\grad}{\ensuremath{{}^{\circ}}\xspace}
\newcommand{\ee}{\ensuremath{e^{+}e^{-}}\xspace}

\newcommand{\mumu}{\ensuremath{{\mu^{+}\mu^{-}}}\xspace}

\renewcommand{\Im}{\ensuremath{\text{Im}}\xspace}
\renewcommand{\Re}{\ensuremath{\text{Re}}\xspace}
\renewcommand{\epsilon}{\varepsilon}

\DeclareMathOperator{\Res}{Res}
\DeclareMathOperator{\Inter}{Int}
\DeclareMathOperator{\QED}{QED}

\begin{document}

\begin{frontmatter}

\title{Measurement of the ratio of the lepton  widths
  $\Gamma_{ee}/\Gamma_{\mu\mu}$ for the $J/\psi$ meson}

\author[binp,nsu]{V.\,M.\,Aulchenko}
\author[binp,nsu]{E.\,M.\,Baldin\corref{cor}}
\cortext[cor]{Corresponding author, e-mail:  E.M.Baldin@inp.nsk.su}
\author[binp]{A.\,K.\,Barladyan}
\author[binp]{A.\,Yu.\,Barnyakov}
\author[binp]{M.\,Yu.\,Barnyakov}
\author[binp]{S.\,E.\,Baru}
\author[binp]{I.\,Yu.\,Basok}
\author[binp]{A.\,M.\,Batrakov} 
\author[binp]{A.\,E.\,Blinov}
\author[binp,nstu]{V.\,E.\,Blinov}
\author[binp,nsu]{A.\,V.\,Bobrov}
\author[binp]{V.\,S.\,Bobrovnikov}
\author[binp,nsu]{A.\,V.\,Bogomyagkov}
\author[binp,nsu]{A.\,E.\,Bondar}
\author[binp]{A.\,R.\,Buzykaev}
\author[binp,nsu]{S.\,I.\,Eidelman}
\author[binp,nsu,nstu]{D.\,N.\,Grigoriev}
\author[binp]{V.\,R.\,Groshev} 
\author[binp]{Yu.\,M.\,Glukhovchenko}
\author[binp]{V.\,V.\,Gulevich}
\author[binp]{D.\,V.\,Gusev}
\author[binp]{S.\,E.\,Karnaev}
\author[binp]{G.\,V.\,Karpov}
\author[binp]{S.\,V.\,Karpov}
\author[binp]{T.\,A.\,Kharlamova}
\author[binp]{V.\,A.\,Kiselev}
\author[binp]{V.\,V.\,Kolmogorov}
\author[binp,nsu]{S.\,A.\,Kononov}
\author[binp]{K.\,Yu.\,Kotov}
\author[binp,nsu]{E.\,A.\,Kravchenko}
\author[binp]{V.\,N.\,Kudryavtsev}
\author[binp,nsu]{V.\,F.\,Kulikov}
\author[binp,nstu]{G.\,Ya.\,Kurkin}
\author[binp,nsu]{E.\,A.\,Kuper}
\author[binp]{I.\,A.\,Kuyanov}
\author[binp,nstu]{E.\,B.\,Levichev}
\author[binp,nsu]{D.\,A.\,Maksimov}
\author[binp]{V.\,M.\,Malyshev}
\author[binp]{A.\,L.\,Maslennikov}
\author[binp,nsu]{O.\,I.\,Meshkov}
\author[binp]{S.\,I.\,Mishnev}
\author[binp,nsu]{I.\,I.\,Morozov}
\author[binp,nsu]{N.\,Yu.\,Muchnoi}
\author[binp]{V.\,V.\,Neufeld}
\author[binp]{S.\,A.\,Nikitin}
\author[binp,nsu]{I.\,B.\,Nikolaev}
\author[binp]{I.\,N.\,Okunev}
\author[binp,nstu]{A.\,P.\,Onuchin}
\author[binp]{S.\,B.\,Oreshkin}
\author[binp,nsu]{I.\,O.\,Orlov}
\author[binp]{A.\,A.\,Osipov}
\author[binp,nstu]{I.\,V.\,Ovtin}
\author[binp]{S.\,V.\,Peleganchuk}
\author[binp,nstu]{S.\,G.\,Pivovarov}
\author[binp]{P.\,A.\,Piminov}
\author[binp]{V.\,V.\,Petrov}
\author[binp]{A.\,O.\,Poluektov}
\author[binp]{V.\,G.\,Prisekin}
\author[binp,nsu]{O.\,L.\,Rezanova}
\author[binp]{A.\,A.\,Ruban}
\author[binp]{V.\,K.\,Sandyrev}
\author[binp]{G.\,A.\,Savinov}
\author[binp]{A.\,G.\,Shamov}
\author[binp]{D.\,N.\,Shatilov}
\author[binp,nsu]{B.\,A.\,Shwartz}
\author[binp]{E.\,A.\,Simonov}
\author[binp]{S.\,V.\,Sinyatkin}
\author[binp]{A.\,N.\,Skrinsky}
\author[binp,nsu]{A.\,V.\,Sokolov}
\author[binp]{A.\,M.\,Sukharev}
\author[binp,nsu]{E.\,V.\,Starostina}
\author[binp,nsu]{A.\,A.\,Talyshev}
\author[binp]{V.\,A.\,Tayursky}
\author[binp,nsu]{V.\,I.\,Telnov}
\author[binp,nsu]{Yu.\,A.\,Tikhonov}
\author[binp,nsu]{K.\,Yu.\,Todyshev}
\author[binp]{G.\,M.\,Tumaikin}
\author[binp]{Yu.\,V.\,Usov}
\author[binp]{A.\,I.\,Vorobiov}
\author[binp]{V.\,N.\,Zhilich}
\author[binp,nsu]{V.\,V.\,Zhulanov}
\author[binp,nsu]{A.\,N.\,Zhuravlev}

\address[binp]{Budker Institute of Nuclear Physics, 11, akademika
  Lavrentieva prospect,  Novosibirsk, 630090, Russia}
\address[nsu]{Novosibirsk State University, 2, Pirogova street,  Novosibirsk, 630090, Russia}
\address[nstu]{Novosibirsk State Technical University, 20, Karl Marx
  prospect,  Novosibirsk, 630092, Russia}

 \begin{abstract}
   The ratio of the electron and muon widths of the \(J/\psi\) meson
   has been measured using direct $J/\psi$ decays in the KEDR experiment at
   the VEPP-4M electron-positron collider. The result
   \[\Gamma_{e^+e^-}(J/\psi)/\Gamma_{\mu^+\mu^-}(J/\psi)=1.0022\pm0.0044\pm0.0048\ (0.65\%)\]
   is  in  good agreement with the lepton universality. The experience 
   collected during this analysis  will be used for 
    \(J/\psi\) lepton width determination with up to 1\% accuracy.
 \end{abstract}
\end{frontmatter}

\section{Introduction}

The lepton width of a  hadronic resonance \(\Gamma_{\ell\ell}\)
describes fundamental properties of the strong interaction
potential~\cite{Brambilla:2010cs}. Comparison of the electron and
muon widths \(\Gamma_{ee}/\Gamma_{\mu\mu}\) allows one to test the
lepton universality and provides information on the models predicting
new forces differentiating between lepton
spe\-cies~\cite{Batell:2011qq}.

Currently the two most precise values of the ratio  of the \(J/\psi\) meson 
lepton widths come from the CLEO results obtained in
2005~\cite{Li:2005uga} and the recent BESIII 
measurement~\cite{Ablikim:2013pqa}. For that analysis both
experiments used the \(\psi(2S)\to J/\psi\pi^+\pi^-\),
\(J/\psi\to\ell^+\ell^-\) decay chain (\(\ell=e,\mu\)).

Our analysis is based on direct $J/\psi$ decays. Its scheme and the 
sources of systematic uncertainties are completely
different from those in the CLEO and BESIII 
measurements.  This analysis continues the work on the lepton width
determination~\cite{Baldin:PLB:2010} but uses an independent statistical sample.
In the future we anticipate  precise measurement of the $J/\psi$ lepton 
width at the 1\% level.

A large resonance cross section provides  high statistics of
$J/\psi$ decays even with a relatively low collider luminosity. The
integrated luminosity collected off resonance gives information
about the QED continuum background. In addition to subtracting the QED
background for calculating the numbers of the $J/\psi\to\ell^+\ell^-$
decays, one has to suppress cosmic ray events, 
$J/\psi$ hadronic decays and take into account the interference
between the resonance $J/\psi\to\ell^+\ell^-$ process and QED background.
The QED backgrounds for the $J/\psi\to e^+e^-$ and
$J/\psi\to\mu^+\mu^-$ processes are fundamentally different due to
Bhabha scattering.

\section{Experiment}

\begin{figure}[t]
  \centering
  \includegraphics[width=\columnwidth]{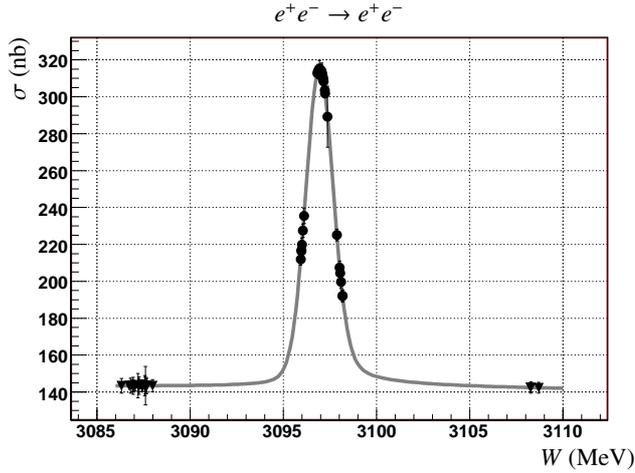}
  \caption{\(e^+e^-\) cross section expected from the
    experimental runs; dots: ``on-resonance'' data, triangles:
    ``off-resonance'' data. The curve is the theoretical di-electron cross
    section~\eqref{eq:ee2ee} in the experimental energy and angle
    ranges.} \label{fig:ee2ee}
\end{figure}

\begin{figure}[t]
  \centering
  \includegraphics[width=\columnwidth]{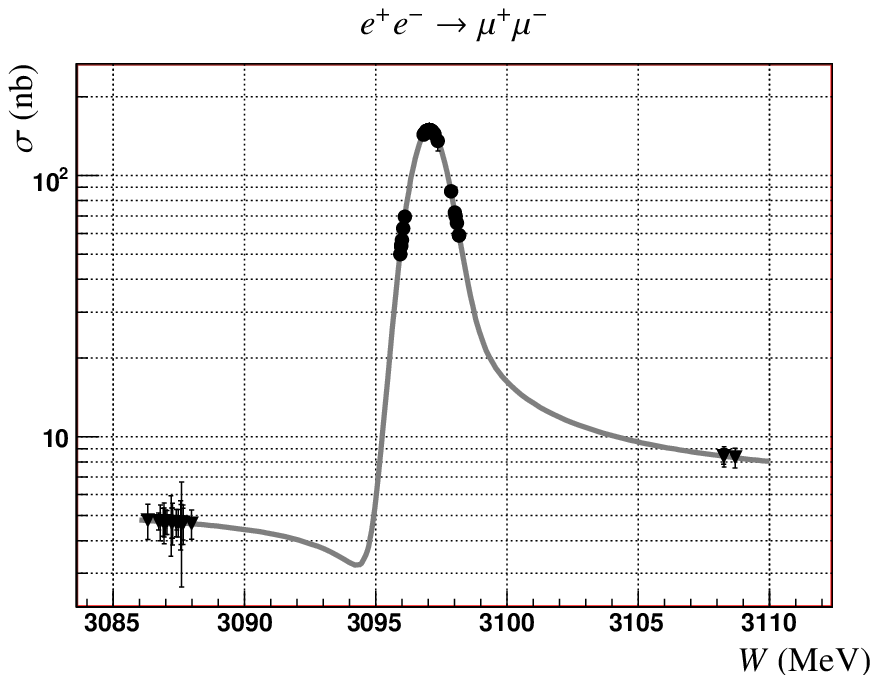}
  \caption{\(\mu^+\mu^-\) cross section expected from  the
    experimental runs; dots:  ``on-resonance'' data, triangles: 
    ``off-resonance'' data. The curve is the theoretical dimuon cross
    section~\eqref{eq:ee2mumu} in the experimental energy and angle
    ranges.} \label{fig:ee2mumu}
\end{figure}

The experiment was performed with the KEDR
detector~\cite{KEDR-detector} at the VEPP-4M \ee
collider~\cite{Anashin:1998sj}. 
An integrated luminosity  of
2.1\,pb$^{-1}$ corresponding to  production  of about 
$6.5\cdot10^{6}$ $J/\psi$ mesons was collected 
in the \(J/\psi\) resonance energy range from 3086 to 3107 MeV. 
The experimental data sample was
divided into two parts (Fig.~\ref{fig:ee2ee} and Fig.~\ref{fig:ee2mumu} for
\(e^+e^-\) and \(\mu^+\mu^-\) events, respectively):  ``on-resonance'', with
\(|W-M_{J/\psi}|<1.3\,\text{MeV}\) (\(\approx 80\%\) of statistics),
and ``off-resonance'', with \(|W-M_{J/\psi}|>8.9\,\text{MeV}\), where
\(W\) is the center-of-mass energy. The energy spread \(\sigma_W\)
was about 0.7\,MeV.

KEDR is a general-purpose detector with solenoidal
 magnetic field. It consists of a vertex detector,
a drift chamber, scintillation time-of-flight counters, 
aerogel Cherenkov counters, a
barrel liquid krypton calorimeter, an endcap CsI calorimeter, and a
muon system built in the yoke of a superconducting coil generating a
field of 0.65 T. The detector also includes a tagging system to
detect scattered electrons for a study of two-photon processes. The
on-line luminosity is measured by two independent
single bremsstrahlung monitors.

The VEPP-4M collider can operate in the wide range of beam energy from
1 to 6 GeV. The peak luminosity in the \(J/\psi\) energy region is
about~\(2\times10^{30}\,\text{cm}^{-2}\text{s}^{-1}\).  One of the
main features of the VEPP-4M is its capability to precisely measure the beam
energy  using two techniques~\cite{Blinov:2009zza}: resonant
depolarization and infrared light Compton backscattering.

\section{Theory}

In the soft photon approximation, analytical expressions for the $e^+e^-\to\ell^+\ell^-$ cross
sections near a  narrow resonance including radiative corrections  are presented in Eqs.~\eqref{eq:ee2ee} and
\eqref{eq:ee2mumu} below:
\begin{equation}\label{eq:ee2ee}
  \begin{split}
    &\left(\frac{d\sigma}{d\Omega}\right)^{ee\to ee}_{\text{th.}}=\frac{1}{M^2}
    \left(1+\delta_{\text{rc}}\right) \Big  \{
    \frac{9}{4}\frac{\Gamma^2_{\ee}}{\Gamma M}
    \left(1+\cos^2\theta\right)\Im\, \mathcal{F} - \\
    &\hspace{0.3em}\frac{3\alpha}{2}\frac{\Gamma_{\ee}}{M}
    \left [(1+\cos^2\theta)\!-\!\frac{(1+\cos\theta)^2}{(1-\cos\theta)}
    \right ]\Re\, \mathcal{F}\Big\}+\left(\frac{d\sigma}{d\Omega}\right)_{\text{QED}}^{ee},
  \end{split}
\end{equation}

\begin{equation}\label{eq:ee2mumu}
  \begin{split}
    &\left(\frac{d\sigma}{d\Omega}\right)^{ee\to\mu\mu}_{\text{th.}}=
    \frac{1}{M^2}\left(1+\delta_{\text{rc}}\right)
    \Big \{\frac{9}{4}\frac{\Gamma_{\ee}\Gamma_{\mumu}}{\Gamma M}
    \Im\, \mathcal{F} - \\
    &\hspace{3em}\frac{3\alpha}{2}\frac{\sqrt{\Gamma_{\ee}\Gamma_{\mumu}}}{M}
    \Re\, \mathcal{F}\Big \}\left(1+\cos^2\theta\right)+
    \left(\frac{d\sigma}{d\Omega}\right)_{\text{QED}}^{\mu\mu},
  \end{split}
\end{equation}
\begin{equation*}
  \begin{split}
    \label{eq:F}
    &\mathcal{F}=
    \frac{\pi\beta}{\sin\left(\pi\beta\right)}\left(\frac{\frac{M}{2}}{-W+M-\frac{i\Gamma}{2}}\right)^{1-\beta}\!\!\!\!\!,\,\,
    \beta=\frac{4\alpha}{\pi}\left(\ln\frac{W}{m_e}-\frac{1}{2}\right)\simeq0.077,\\
    &\delta_{\text{rc}}=\frac{3}{4}\beta+
    \frac{\alpha}{\pi}\left(\frac{\pi^2}{3}-\frac{1}{2}\right)+
    \beta^2\left(\frac{37}{96}-\frac{\pi^2}{12}-\frac{\ln(W/m_e)}{36}\right),
  \end{split}
\end{equation*}
where \(W\) is the center-of-mass energy, and \(\theta\) is the lepton
scattering angle with respect  to the electron beam direction.  
Corrections to the
vacuum polarization in the interference terms have been omitted.

The formulae used in this analysis are based on the analytical
expression for the radiative correction integral in the soft photon
approximation (SPA), first obtained in~\cite{azimov-1975-eng}.  The
accuracy was improved using~\cite{KuraevFadin} as described
in~\cite{Anashin:2011kq}.

To compare experimental data with the theoretical cross sections, it
is necessary to perform their convolution with a distribution of the
total beam energy, which is assumed to be Gaussian:
\begin{equation}
  \label{eq:1}
  \sigma^{\ell\ell}(W)=\int\frac{1}{\sqrt{2\pi}\,\sigma_W}
  \exp\left(-\frac{(W-W')^2}{2\sigma^2_W}\right)
  \sigma^{\ell\ell}_{\text{th.}}(W')\,dW'.
\end{equation}
The beam energy spread \(\sigma_W\) is much larger than the \(J/\psi\)
full width \(\Gamma\).

\section{Event selection}

The following selection requirements were imposed on both \(e^+e^-\)
and \(\mu^+\mu^-\) events (the $+$ and $-$ superscripts  correspond to a
positive particle  and negative one, respectively):
\begin{enumerate}
\item two charged tracks with opposite signs from a common vertex in
  the interaction region,
\item the total energy deposition in the calorimeter (outside the two
energy clusters belonging to the selected particles) is \(<0.15\,\text{GeV}\),
\item polar \(\theta\) and azimuthal \(\varphi\) acollinearity \(<10\grad\),
\item the momentum \(p^{\pm}>0.5\)\,GeV.
\end{enumerate}

Only for the \(e^+e^-\) selection: the energy deposition for each particle
$E^{\pm}>0.7$\,GeV; \(\theta^{-}\in (41\div139)\grad\) and
\(\theta^{+}\in (38\div142)\grad\).  The fiducial polar angle
\(\theta\) is restricted by the  physical edges of the liquid krypton
calorimeter $(37\div143)\grad$.

Only for the \(\mu^+\mu^-\) selection:
$0.06\,\text{GeV}<E^{\pm}<0.7\,\text{GeV}$; \(\theta^{-}\in
(49\div131)\grad\) and \(\theta^{+}\in (46\div134)\grad\).  The polar
angle \(\theta\) is restricted by the edges of the muon system.  To
suppress the background of cosmic events we employed the
time-of-flight system.  For suppression of the background from
\(J/\psi\) hadronic decays, a hit in the muon system is required for a
\(\mu^-\) track.

\begin{figure}[t]
  \includegraphics[width=\columnwidth]{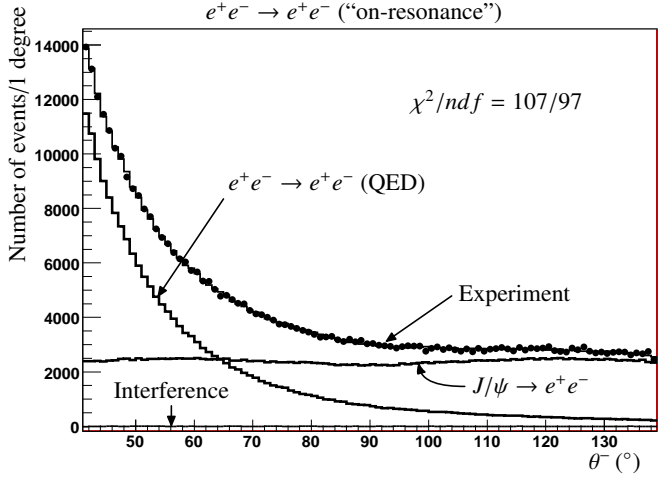}
  \caption{Distribution of the electron scattering angle for selected
    $e^+e^-\to e^+e^-$ events in the ``on-resonance'' data part.  The
    dots represent the experiment. The labeled histograms represent
    the simulation: Bhabha (QED), $J/\psi\to e^+e^-$ decays and
    interference term.  The  histogram  corresponding to the sum of
    the three contributions is under the experimental dots.} \label{fig:hpee}

\end{figure}

\begin{figure}[t]
  \includegraphics[width=\columnwidth]{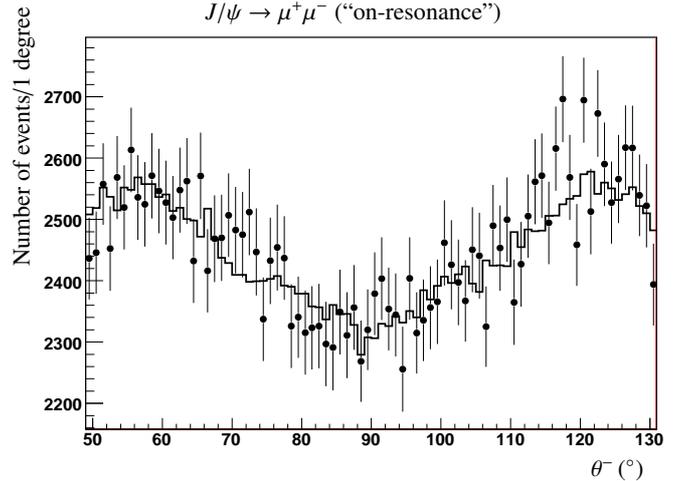}
  \caption{Distribution of the \(\mu^-\) scattering angle for
    $J/\psi\to \mu^+\mu^-$ events in the ``on-resonance'' data part.
    The dots represent the experiment. The histogram represents the
    simulation.} \label{fig:hpmumu}

\end{figure}

The \(\theta^-\) angle distributions of the selected events are shown in
Fig.~\ref{fig:hpee} and Fig.~\ref{fig:hpmumu} for \(e^+e^-\) and
\(\mu^+\mu^-\), respectively.

\section{Simulation}
\label{sec:simulation}

The contributions of the \ee and \mumu resonance and interference
events were simulated according to the theoretical angular
distributions~\eqref{eq:ee2ee} and \eqref{eq:ee2mumu}, respectively.
The final state radiation was taken into account using the
PHOTOS~\cite{photos} package.

For the resonance contribution of \mumu due to the relatively high mass of
muons, a more precise expression for the angular distribution
\(\frac{d\sigma}{d\Omega}\propto\beta\times\left(1+\cos^2\theta+\left(1-\beta^2\right)\times\sin^2\theta\right)\)
was taken. The
contribution of  the \(\sin^2\theta\) term to this analysis is about 0.2\%.

The uncertainty in the Bhabha process simulation was evaluated via
comparison of our result with two independent event generators:
BHWIDE~\cite{BHWIDE} and MCGPJ~\cite{Arbuzov:2005pt}. For
the \(\ee\to\gamma\gamma\) process, the BABAYAGA
generator~\cite{CarloniCalame:2001ny} was employed. For estimating
the \(J/\psi\) background, the BES generator~\cite{Chen:2000tv} was used.

To take the coincidence of the signal and background into account, we
added the random events recorded every \(5\cdot10^{-5}\) beam crossing
to the simulated events.

\section{$J/\psi\to  \ell^+\ell^-$ event counting}

We begin our analysis by determining independently the number of
\ee and \mumu events produced in $J/\psi$ decays.

\begin{table}[t]
\begin{tabular}{lccc}\hline
Event number  & \(N_{ee}\)& \(N_{\mu\mu}\)\\\hline
 \quad ``on-resonance'' &\(425786\pm658\) & \(162515\pm406\)\\ 
 \quad QED (bck)&\(190345\pm770\) & \(5750\pm181\)\\ 
 \quad \(J/\psi\to\text{hadr.}\) (bck) &\(373\) & \(215\)\\
 \quad \(J/\psi\to\ell\ell\) decays &\(235298\pm774\)&\(156550\pm447\) \\\hline 
 Efficiency  & \(\varepsilon_{ee}\), \%& \(\varepsilon_{\mu\mu}\), \%\\\hline
 \quad\(\varepsilon_{\text{ToF}}\)    & --- & 77.78 \\ 
 \quad\(\varepsilon\) & 60.14 & 51.55 \\ \hline
 \(N(J/\psi\to\ell\ell)/\varepsilon\) &\(391281\pm1287\) & \(390412\pm1113\)  \\\hline
\end{tabular}
\caption{Summary of observed events, principal backgrounds,  signals and
  their efficiency. 
  The QED background events also include 
  interference corrections.  The efficiency \(\varepsilon_{\text{ToF}}\)
  corresponds to the ToF time measurement inefficiency. The stated errors
  are statistical only.}
  \label{tab:summary}
\end{table}

A summary of observed events,  principal backgrounds,  signals and
their efficiency is presented in Table~\ref{tab:summary}.

In Fig.~\ref{fig:hpee} we show the distribution of the electron
scattering angle for selected $e^+e^-\to e^+e^-$ events in the
resonance data part.  The displayed points represent the experimental
values, while the histograms correspond to the simulation. The 
angular distribution of  Bhabha events
differs from that of \(J/\psi\to e^+e^-\) decays.
At small angles the Bhabha scattering prevails, while at large angles
events of resonance decay dominate.  So, these processes can be
separated by using only a data sample collected 
``on-resonance''. The ``off-resonance'' events are not required. 

The ``on-resonance'' data sample was collected in the vicinity of the
resonance peak.  Thus we have to take into account the interference
effects between $J/\psi$ decays and QED background (the
histogram close to zero in Fig.~\ref{fig:hpee}). However, the
interference effects are an \(\sim 1\%\) correction only.

For separating \(J/\psi\to\ee\) events from the Bhabha QED background,
the number of observed experimental events was fitted to the
expected contributions:
\begin{equation}
  \frac{dN^{\text{obs}}_{ee}(\theta)}{d\theta}=n\times\Res(\theta)+
   \left<C(E)\right>\times
    \Inter(\theta)+
  L\times\QED(\theta),\label{eq:Neeobs}   
\end{equation}
where \(n\) and \(L\) are the fit parameters which correspond to the
number of observed \(J/\psi\to\ee\) events and to the absolute
luminosity calibration, respectively.
\(\Res(\theta)\), \(\Inter(\theta)\) and \(\QED(\theta)\) are the
angular distributions from the simulation for the resonance,
interference and Bhabha QED background, respectively
(sec.~\ref{sec:simulation}).  The same histograms as presented in
Fig.~\ref{fig:hpee} with one degree bin width were used in the fitting 
procedure.
Thus from the detection efficiency $\varepsilon_{J/\psi\to ee}$ we can
calculate the number of \(J/\psi\) decays during the experiment:
\(N_{J/\psi\to ee}=n/\varepsilon_{J/\psi\to ee}\).  The \(\Res(\theta)\),
\(\Inter(\theta)\) and \(\QED(\theta)\) angular distributions and this
efficiency were determined from the simulation and corrected using
information about performance of various detector subsystems.  The
statistical uncertainty of the number of \ee decays is 0.33\%.

The \(\left<C(E)\right>\) coefficient, which reflects the energy
variation in the data set, is calculated from theory,
see Eq.~\eqref{eq:ee2ee}, and is determined by the interference magnitude.

The same procedure was performed for the continuum statistics,
since  in our ``off-resonance'' data, the resonant
contribution and interference effects are also not completely negligible.
The number of Bhabha events in continuum is necessary to calculate the
number of \(J/\psi\to\mu^+\mu^-\) decays.

For calculating the number of \(\mu^+\mu^-\) decays
(Fig.~\ref{fig:ee2mumu}) we have to take the interference into account,
subtract the QED background and divide that by the detection efficiency:
\begin{equation}
N_{J/\psi\to\mu\mu}=
 \frac{\left\{N^{\text{exp}}_{\text{res\phantom{t}}}-N^{\text{th}}_{\text{int}}-
   \frac{L_{\text{res\phantom{y}}}}{L_{\text{cont}}}\times
 \left(N^{\text{exp}}_{\text{cont}}-N^{\text{th}}_{\text{int}}\right)\right\}}
 {\varepsilon_{J/\psi\to\mu\mu}}.
 \label{eq:Nmumu}
\end{equation}

As in the $e^+e^-$ case, the efficiency was determined from the
simulation and corrected using information about performance of
various detector subsystems. The statistical error of the number of
\(J/\psi\to\mumu\) decays is 0.29\%.

\section{Systematic uncertainties}

\begin{table}[t]
  \centering
  \begin{tabular}{p{0.45\columnwidth}rr}\hline
    Source  & Correct.,\,\% & Err.,\, \%  \\\hline
    Interference   &    &       \\
    \quad Luminosity & &\(0.01\)\\
    \quad Energy measurement   &    &\(0.02\)\\
    \quad Radiation corrections &  & \(0.10\) \\
    Background  &   &\\
    \quad$J/\psi\to\text{hadrons}$ & \(-0.05\) &\(0.10\) \\
    \quad $\ee\to\gamma\gamma$  & \(-0.07\)  & \\
    \quad Cosmic     & &\(0.07\)   \\
    Simulation       &          &           \\
    \quad Bhabha       & & \(0.11\)\\
    \quad PHOTOS       &\(+0.20\) &\(0.02\)\\
    Trigger           &             &          \\
    \quad 1st level   &\(-0.70\)  &\(0.20\)\\
    \quad 2nd level   &\(-1.17\)  &\(0.11\)\\
    Event selection           &         &           \\
    \quad tracking system  & \(+1.18\) & \(0.10\)\\
    \quad calorimeter     & \(+0.27\)   & \(0.10\)\\
    \quad muon system  &\(-0.12\) &\(0.04\) \\
    \quad \(\theta\) angle  cuts && \(0.10\)\\
    \(\theta\) angle determination & & \(0.14\)\\
    Selection asymmetry  & & \(0.14\)\\
    ToF inefficiency  &\(-22.2\)  &\(0.26\) \\\hline
    \multicolumn{2}{l}{\emph{Total Systematic Uncertainties}} &\emph{0.48}\\\hline
  \end{tabular}
  \caption{Summary of the systematic relative uncertainties for the ratio of
    \(\Gamma_{ee}/\Gamma_{\mu\mu}\). The first column is the source of 
    uncertainty, the second is the correction to the result which was
    applied due to this source if applicable and the third one is the
    uncertainty.}
  \label{tab:sys}
\end{table}

A list of main systematic uncertainties in the ratio of
\(\Gamma_{ee}/\Gamma_{\mu\mu}\) is presented in Table~\ref{tab:sys}.

Luminosity, energy measurement and theoretical radiation
corrections 
are important mainly for the interference effects, which are
small corrections only.

The  $J/\psi$ hadronic decay contribution to the selected \mumu and \ee events
was estimated by the Monte Carlo method and
the scale of uncertainty was estimated using the nuclear
interaction simulation packages FLUKA~\cite{FLUKA1} and
GHEISHA~\cite{Fesefeldt:1985yw} (as implemented in
GEANT~3.21~\cite{GEANT-www}). The uncertainty of the contribution from the
\(\ee\to\gamma\gamma\) background is negligible. A possible
contribution to \ee events from cosmic events was estimated using the
muon and the time-of-flight systems.

From the trigger ``point of view'', the main difference between \ee
and \mumu events is the high energy deposition for \ee events in the
calorimeter.  The trigger calorimeter thresholds were too high for
\mumu events, but the efficiency for \ee selected events is
99.0\%.  The efficiency of the calorimeter trigger for
  \ee selected events was estimated with the help of the ToF system
  trigger. The first level trigger selected \mumu events using the
ToF system only, but \ee events in addition to the ToF system, could
be independently selected with the calorimeter.

The first level trigger \mumu inefficiency \mbox{\(\simeq0.7\%\)} was
measured using the data from special ``cosmic'' runs with soft ToF
system restrictions. The uncertainty \(\simeq0.2\%\) was estimated by
comparison of the inefficiency obtained with differently selected
subsets of the cosmic events.  The second level trigger \mumu
inefficiency \(\simeq1.17\%\) and uncertainty \(\simeq0.11\%\), mainly
due to the vertex detector, were estimated using the \ee data.

The corrections to the detector efficiency were obtained using
experimental data. Event selection uncertainties were estimated via 
variations of
the cuts.  The uncertainty of the \(\theta\) angle determination was
evaluated via comparison of the angular measurements 
performed in  the tracking system and the liquid krypton calorimeter.

The event selection was asymmetrical with respect to the particle
sign.  The same procedures were performed with the opposite sign. The
final result is the half-sum and the estimated uncertainty is the
half-difference of the results of these two procedures.

\begin{figure}[t]
  \centering
  \includegraphics[width=\columnwidth]{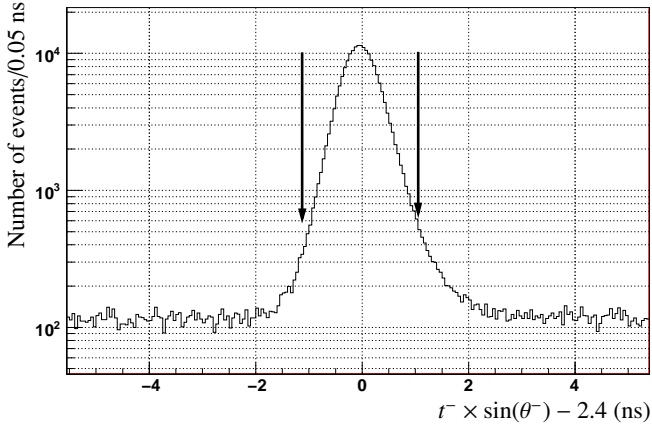}
  \caption{The ToF time distribution for \(\mu^-\). The arrows
    represent the ToF cuts \(\pm3\sigma_t\), where
    \(\sigma_t=0.36\)\,ns is a ToF time resolution.}
  \label{fig:tofst}
\end{figure}

The main error comes from the ToF time measurement inefficiency due to
a dead time in the time expander.  It is a rather large correction as
compared with the others.  The time distribution for \(\mu^-\) is
presented in Fig.~\ref{fig:tofst}.  The cosmic background is flat and
could be easily measured. Thus it is possible to estimate the
efficiency for the \(\mu^+\) with \(\mu^-\) time cuts applied and vise
versa.  The net time-of-flight efficiency is a product of these
values:
\[\epsilon_{\text{ToF}}=\varepsilon_{\mu^+
  (\mu^-)}\times\varepsilon_{\mu^- (\mu^+)}=(77.78\pm0.12\pm0.03)\%\]
provided that there are no correlations.  This assumption was checked
with an electron data sample (\(\varepsilon^{ee}_{\text{ToF
    real}}=76.35\%\) as compared with \(\varepsilon^{ee}_{\text{ToF
    calc}}=\varepsilon_{e^+(e^-)}\times\varepsilon_{e^-(e^+)}=76.51\%\)).
The relative difference 
\[
\delta\varepsilon_{\text{ToF}}/\varepsilon_{\text{ToF}}=
(\varepsilon^{ee}_{\text{ToF real}}- 
 \varepsilon^{ee}_{\text{ToF calc}})/
\varepsilon^{ee}_{\text{ToF}}=0.21\%\] 
estimates the possible correlation magnitude.  Adding the statistical
uncertainty (\(\simeq0.12\%\)) and uncertainty from the cosmic
background (\(\simeq0.03\%\)) estimation in quadrature, we obtain a
total systematic uncertainty of 0.26\%.

\section{Result}

\begin{figure}[t]
  \centering
  \includegraphics[width=0.8\columnwidth]{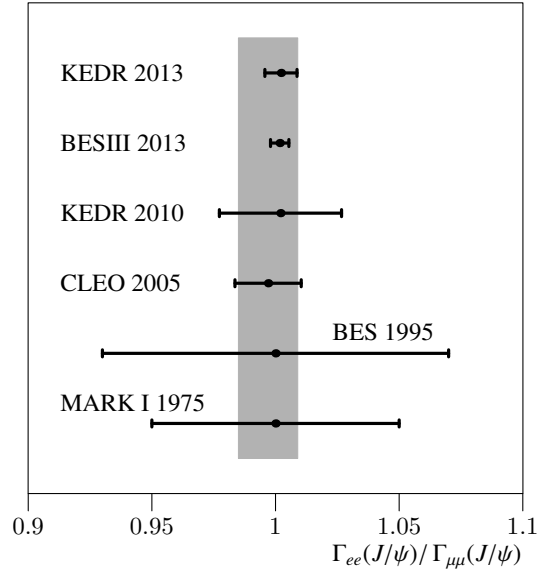}  
  \caption{Comparison of the current and earlier measurements of the ratio
$\Gamma_{ee}/\Gamma_{\mu\mu}$. The vertical grey band marks the
average value and uncertainty of the 2012 PDG compilation~\cite{PDG-2012}.}
\label{fig:GeeVsGmumu}
\end{figure}

To conclude, a measurement of the ratio of the lepton widths
$\Gamma_{ee}$ and $\Gamma_{\mu\mu}$
has been performed at the
VEPP-4M collider using the KEDR detector. 
Our final result is as follows:
\[\Gamma_{e^+e^-}(J/\psi)/\Gamma_{\mu^+\mu^-}(J/\psi)=1.0022\pm0.0044\pm0.0048.\]
Adding the statistical and systematic errors in quadrature, we obtain a
total ratio uncertainty of about 0.65\%.  This result is in good agreement with
the lepton universality and provides information for the models predicting new 
forces
 differentiating between lepton species~\cite{Batell:2011qq}.   
A comparison with other measurements is
presented in Fig.~\ref{fig:GeeVsGmumu}. The experience collected during
this analysis will be used for a  measurement of the $J/\psi$ lepton
width at the 1\% level, important for various applications, e.g.,
for a determination of the charm quark mass~\cite{Kuhn:2007vp}.  
 
The authors are grateful to J. K\"{u}hn and M. Pospelov for fruitful
discussions.
This work is supported by the Ministry of Education and Science of the
Russian Federation, RFBR grants \mbox{12-02-00023-a}, 12-02-01076-a,
14-02-31401,  Sci. School Nsh-5320.2012.2 grant and  DFG grant HA
1457/9-1.

\end{document}